\documentstyle[12pt]{article}
\newcommand{\be}{\begin{equation}}
\newcommand{\ee}{\end{equation}}
\newcommand{\ed}{\end{document}}
\newcommand{\lab}[1]{\label{#1}}
\newcommand{\re}[1]{(\ref{#1})}
\newcommand{\ci}[1]{\cite{#1}}
\renewcommand{\baselinestretch}{1.4}
\date{}
\title{ STOCHASTIC IONIZATION OF RELATIVISTIC HYDROGEN-LIKE ATOM }
\author{ D.U.MATRASULOV \\
 Heat Physics Department of the Uzbek Academy of Sciences,\\
28 Katartal St.,700135 Tashkent, Uzbekistan \thanks{e-mail:
davron@silk.org }}
\begin{document}\large
\begin{titlepage}
\maketitle

\begin{abstract}
Stochastic ionization of higly excited relativistic hydrogenlike
atom in the monochromatic field is considered. A theoretical analisis
of chaotic dynamics of the electron based on Chirikov criterion is given.
Critical value of the external field is evaluated analitically.
\end{abstract}
\end{titlepage}

\section*{Introduction}

Study of behaviour of higly exited atoms in microwave field is important
problem which lies at the intersection of several lines of contemporary
research. On of important of this themes is chaos. Recently chaotic
dynamics of atom in the microwave field have been subject of extensive
reseach. Much theoretical and experimental work was devoted to this
problem \ci{1,2,3}. A theoretical analisis of behaviour of classical
hydrogen atom, based on Chirikov criterion \ci{3,4} shows that  for some
critical value of field strength, the electron enters a chaotic regime
of motion, marked by unlimited diffusion leading to ionization.
Up to now much of the discussion on resonance overlap and chaos has
largely been limited by nonrelativistic systems.
 Investigation of classical relativistic systems with chaotic dynamics
has became of subject of extensive research in the last years
\ci{11}-\ci{14}. Among these considered problems are some nonlinear oscillators\ci{11} relativistic
two-center Kepler problem\ci{12}, etc.
In our opinion, the study of chaotic dynamics of relativistic atoms
interacting with microwave field makes possible to investigate the new
phenomenon, namely, relativistic quantum chaos.

In the present paper we generalize
mentioned above results on chaotic ionization of classical nonrelativistic
hydrogen atom to the case of classical relativistic hydrogen-like atom.
For simplicity we consider the one-dimensional model which allows to estimate
analitically the values of critical field strength. As is well known (see \ci{8})
and references therein, in the nonrelativistic case one of the remarkable futures
of the comparison of the classical theory with the experimental measurements
of microwave ionization is the fact that a one-dimensional model of a hydrogen
atom in an oscillating electric field provides an excellent description of
experimental ionization thresholds for real three-dimensional hydrogen atom.
In the case of relativistic hydrogen-like atom one-dimensional model allows to
avoid the difficulties connected with the non-closeness of trajectories in the
relativistic Kepler motion \ci{15} and arising additional degrees of freedom
\ci{16}.

Using
Chirikov's criterion of stochasticity we obtain, in terms of action and charge,
the analitical formula for the critical value of microwave field, at which stochastic ionization
will occur.
In this paper we will use the system of units $m_{e} =h= c
=1 $ in which $e^2 = 137^{-1}$.

Consider classical relativistic electron in a field of
one-dimensional $-\frac{Z\alpha} {x}$ potential
 where $Z$ is the charge of the center, $\alpha =
\frac{1}{137}$. Momentum of this electron is given by $$ p =
 \sqrt{(\varepsilon +\frac{Z\alpha}{x})^2 -1}, $$ where $\varepsilon$ is
the full energy of the electron.  Action is defined as $$ n =
  \int^{x_{1}}_{x_{2}}p dx  = \pi a\sqrt{1-\varepsilon^2}, $$ where $ a
 =\frac{\varepsilon Z\alpha}{1-\varepsilon^2}$ $x_1$ and $x_2$ are the
turning points of the electron. From this expression  we have for
unperturbed Hamiltonian \be H_{0} = \frac{n}{\sqrt{n^2 +\pi^2
Z^2\alpha^2}} \lab{unpert} \ee The corresponding frequency is
\be
\omega_{0} = \frac{dH_{0}}{dn} =
\frac{\pi^2Z^2\alpha^2}{(n^2+\pi^2Z^2\alpha^2)^{\frac{3}{2}}}
\lab{freq}
\ee
It is easy to see that in the nonrelativistic limit (i.e., for small Z)
both Hamiltonin and frequency coincide with corresponding nonrelativistic
ones \ci{1,5}. Now we consider interaction of our atom with monchromatic
field which is written in the form
\be
V(x,t) = \epsilon xcos\omega t ,
\lab{potential}
\ee
where $\epsilon$ and $\omega$ are the field amplitude and frequency.
First we need to write \re{potential} in the action-angle variables. This
can be done by expanding perturbation \re{potential} into Fourier series:
\be
V(x,t) = \epsilon \sum^{\infty}_{-\infty} x_{k}(n)cos(k\theta - \omega t),
\lab{Fourier}
\ee
where Fourier amplitudes of the perturbation are defined by the integral
\be
x_{k} = \int^{2\pi}_{0}d\theta e^{im\theta}x(\theta,n) =
-\frac{a}{k}J'(ek) = -\frac{n\sqrt{n^2 +Z^2}}{k}J'(ek),
\ee
$J'_{k}$ is the ordinary Bessel function of order k,
$e = \frac{\sqrt{n^2 +Z^2}}{n}$.

Thus full Hamiltonian of the relativistic hydrogenlike atom in the
monochromotic field can be written in the form
\be
H= \frac{n}{\sqrt{n^2 +\pi^2 Z^2\alpha^2}} +
 \epsilon \sum^{\infty}_{-\infty} x_{k}(n)cos(k\theta - \omega t).
\lab{full}
\ee

For sufficiently small electric fields the Kolmogorov-Arnol'd-Moser
theorem \ci{10} guarantes that most of the straight-line trajectories
in action-angle space will be only slightly distorted by the perturbation.
The maximum distortion of the orbits will occur at resonances where the
phase, $k\theta - \omega t $, is stationary \ci{10}. The resonance
frequency and actions are threrefore detaermined by the relation
\be
k\omega_{0} - \omega = 0
\lab{resonance}
\ee
Then using Eqs. \re{freq} and \re{resonance} the action resonant with the
$k$th subharmonic of the perturbation is
\be
n_{k} = [(\frac{\pi^2
kZ^2\alpha^2}{\omega^2})^{\frac{2}{3}} - \pi^2 Z^2\alpha^2 ]^{\frac{1}{2}}
\lab{action}
\ee

 As is well known there are several methods for investigating of
chaotical dynamicsof the system with Hamiltonian in the form \re{full}.
Simplest of them is one, developed by Zaslavsky and Chirikov  \ci{4,6,7}
which is called Chirikov criterion. According to this criterion chaotical
motion occurs when two neighbouring resonances overlap i.e., when the
following condition is obeyed :
\be \frac{\Delta n_{k}}{\delta n_{k}} > 1 ,
\lab{overlap}
\ee
where $\Delta n_{k}$ is the width of $k$ th resonace,
$\delta n_{k} = n_{k+1} - n_{k}$ is the separation of the $k$  and $k+1$
resonances. Using \re{action} this separation is
$$
\delta n_{k} = (\frac{kZ^2}{\omega})^{\frac{2}{3}}\frac{1}{3kn_{k}}
$$
According to \ci{7,10} the width of $k$th resonance defined as
\be
\Delta n_{k} = 4(\frac{\epsilon x_{k}}{\omega'_{0}})^{\frac{1}{2}}
\ee
where
$$ \omega'_{0} = \frac{d\omega_{0}}{dn} = \frac{3n_{k}Z^2}
{(n_{k}^2 +Z^2)^{\frac{5}{2}}}. $$
Note that these formulas for $\Delta n_{k}$ and $\delta n_{k}$ in the
limit of small $Z$ coincides for known formulas corresponding to
nonrelativistic case \ci{1}.

Taking into account expression for $a$ and using the asymptotic formula
$ J'_{k}(ek) \approx 0.411k^{-\frac{5}{3}}$ (for $k>>1$) for the
resonance width we obtain
$$
\Delta n_{k} \approx [\epsilon
k^{-\frac{8}{3}}Z^{-3}(n_{k}^2 +\pi^2Z^2\alpha^2]^{\frac{1}{2}}$$
Inserting expressions for $\Delta n_{k}$ and $\delta n_{k}$ into
\re{overlap} we have
$$
\epsilon^{\frac{1}{2}}k^{-\frac{1}{3}}Z^{-\frac{3}{2}}n_{k}
(n_{k}^2 +\pi^2 Z^2\alpha^2)^{\frac{1}{2}} >1
$$
This gives us the critical value of the field amplitude at which
sotchastic ionization of the relativistic electron binding in the
field of charge $Z\alpha$ will occur:
$$
\epsilon_{cr} = k^{\frac{2}{3}}(\pi Z\alpha)^3 n_{k}^{-2}(n_{k}^2 +
\pi^2 Z^2\alpha^2)^{-1}
$$
For small $Z$ last formula can be expanded into series as follow:
$$
\epsilon_{cr} = k^{\frac{2}{3}}Z^3n_{k}^{-4}(1 - \frac{Z^2}{n_{k}^2}
+ ... )
$$
or
$$
\epsilon_{cr} = \epsilon_{nonrel}(1 - \frac{Z^2}{n_{k}^2}
+ ... )
$$
where  $\epsilon_{nonrel}$ is the critical field corresponding to the
nonrelativistic case.

As seen from this formula the critical field requiring for stochastic
ionization of relativistic hydrogen-like atom is less than the
corresponding nonrelativistic one.

We have obtained approximate analitical formula for the critical value
of the field requiring for stochastic ionization of relativistic electron
binding in the Coulomb field of charge $Z$ in terms of $Z$ and action
$n_{k}$. Since relativistic Rydberg atom is an esentially quantum object,
the study of its microwave field excitation, provides, therefore an ideal
testing ground for the existence of quantum relativictic "chaotic"
phenomena. More detail analisis of considered above problem should be
given by solving the time-dependent Dirac equation and classical equations
of motion. This will be subject of our future publications.

\ed